\documentclass[prl,twocolumn,preprintnumbers,amsmath,amssymb,superscriptaddress,floatfix,showpacs]{revtex4}
\usepackage{bm}
\usepackage[final]{graphicx}
\usepackage{epsfig}

\begin{document}
\title{Collective spin excitations in a quantum spin ladder probed by high-resolution Resonant Inelastic X-ray Scattering}
\date{\today}
\author{J. Schlappa}
\affiliation{Paul Scherrer Institut, CH-5232 Villigen PSI,
Switzerland}
\author{T. Schmitt} \email{thorsten.schmitt@psi.ch}
\affiliation{Paul Scherrer Institut, CH-5232 Villigen PSI,
Switzerland}
\author{F. Vernay}
\affiliation{Paul Scherrer Institut, CH-5232 Villigen PSI,
Switzerland}
\author{V.~N. Strocov}
\affiliation{Paul Scherrer Institut, CH-5232 Villigen PSI,
Switzerland}
\author{V. Ilakovac}
\affiliation{Universit\'{e} Pierre et Marie Curie - CNRS UMR 7614,
LCP-MR, Paris, France} \affiliation{Universit\'{e} de
Cergy-Pontoise, D\'{e}partement de Physique, F-95000 Cergy-Pontoise,
France}
\author{B. Thielemann}
\affiliation{Laboratory for Neutron Scattering, ETH Zurich and Paul
Scherrer Institut, CH-5232 Villigen PSI, Switzerland}
\author{H.~M. R{\o}nnow}
\affiliation{Ecole Polytechnique F\'{e}d\'{e}rale de Lausanne
(EPFL), CH-1015 Lausanne, Switzerland}
\author{Vanishri S.}
\affiliation{INAC/SPSMS/DRFMC, CEA-Grenoble, 17, rue des Martyrs,
38054 Grenoble Cedex 9, France}
\author{A. Piazzalunga}
\affiliation{CNR/INFM Coherentia/Soft - Dip. Fisica, Politecnico di
Milano, p. Leonardo da Vinci 32, 20133 Milano, Italy}
\author{X. Wang}
\affiliation{Ecole Polytechnique F\'{e}d\'{e}rale de Lausanne
(EPFL), CH-1015 Lausanne, Switzerland}
\author{L. Braicovich}
\affiliation{CNR/INFM Coherentia/Soft - Dip. Fisica, Politecnico di
Milano, p. Leonardo da Vinci 32, 20133 Milano, Italy}
\author{G. Ghiringhelli}
\affiliation{CNR/INFM Coherentia/Soft - Dip. Fisica, Politecnico di
Milano, p. Leonardo da Vinci 32, 20133 Milano, Italy}
\author{C. Marin}
\affiliation{INAC/SPSMS/DRFMC, CEA-Grenoble, 17, rue des Martyrs,
38054 Grenoble Cedex 9, France}
\author{J. Mesot}
\affiliation{Laboratory for Neutron Scattering, ETH Zurich and Paul
Scherrer Institut, CH-5232 Villigen PSI, Switzerland}
\affiliation{Ecole Polytechnique F\'{e}d\'{e}rale de Lausanne
(EPFL), CH-1015 Lausanne, Switzerland}
\author{B. Delley}
\affiliation{Paul Scherrer Institut, CH-5232 Villigen PSI,
Switzerland}
\author{L. Patthey}
\affiliation{Paul Scherrer Institut, CH-5232 Villigen PSI,
Switzerland}
\begin{abstract}
We investigate magnetic excitations in the spin-ladder compound
Sr$_{14}$Cu$_{24}$O$_{41}$ using high-resolution Cu $L_3$-edge
Resonant Inelastic X-ray Scattering (RIXS). Our findings demonstrate
that RIXS couples to collective spin excitations from a quantum
spin-liquid ground state. In contrast to Inelastic Neutron
Scattering (INS), the RIXS cross section changes only moderately
over the entire Brillouin Zone (BZ), revealing a high sensitivity
also at small momentum transfers. The two-triplon energy gap is
found to be $100\pm 30$~meV. Our results are supported by
calculations within an effective Hubbard model for a finite-size
cluster.
\end{abstract}
\maketitle

Collective excitations in strongly correlated electron materials
remain a pivotal challenge in contemporary solid state physics.
Magnetic excitations are heavily debated to provide the pairing
interaction in the high-temperature and unconventional
superconductors \cite{Mathur1998,Monthoux2007}. From that
perspective quantum spin systems attract considerable interest.
While most such materials, e.g., the cuprate superconductors,
exhibit enormous complexity, the two-leg spin ladder is easier to
tract theoretically
\cite{Dagotto1992,Rice1993,DagottoRice1996,Eccleston1998}. It
consists of two parallel chains (legs) with a transverse (rung)
exchange coupling. This system features a singlet ground state and
dispersive triplet excitations, that both have quantum mechanical
origin without any classical counterpart. To date, mainly two
techniques have been established as momentum- and energy-resolved
probes of the dispersion of collective excitations: angle-resolved
photoelectron spectroscopy (ARPES) and inelastic neutron scattering
(INS) for charge and spin degrees of freedom, respectively
\cite{Damascelli2003,Tranquada_fromBook}. Due to the latest
instrumental improvements \cite{GhiringhelliRSI,Strokov}, the energy
scale of magnetic exchange is becoming readily accessible for
resonant inelastic x-ray scattering (RIXS)
\cite{Kotani2001,Harada2002,Chiuzbaian2005,Duda2006}, which is
promising to give information on both, spin and charge degrees of
freedom, and in addition is an element-specific technique.
Furthermore, RIXS requires only small sample volumes
($<0.1$~mm$^3$). Recent RIXS studies of magnetic systems focussed on
spin excitations in long-range ordered magnets
\cite{Ghiringhelli2008,Hill2008,Braicovich2008}.

In this letter, we report our study of the two-leg quantum spin
ladder Sr$_{14}$Cu$_{24}$O$_{41}$ \cite{McCarron1988,Vuletic2006} by
means of momentum-resolved high-resolution RIXS at the Cu $L_3$
edge. One important question is how RIXS, which also couples to
charge, can provide information on magnetic excitations from such a
quantum ground state. We demonstrate unambigously that this
technique couples to purely quantum mechanical excitations from a
singlet ground state. While the INS cross section is inherently low
around the Brillouin zone (BZ) center (small momentum transfers) in
a similar compound \cite{Notbohm2007}, the observed RIXS signal is
found to be intense all over the BZ. A numerical investigation of a
Hubbard model as well as the optical transition selection rules
leads us to conclude that the response is due to two-triplon
excitations. We demonstrate that in the case of a gapped spin
liquid, RIXS is particularly sensitive to these excitations. We thus
are in position to directly evaluate the two-triplon energy gap in
Sr$_{14}$Cu$_{24}$O$_{41}$ at zero momentum transfer as
$100\pm30$~meV.

RIXS experiments were performed at the Advanced Resonant
Spectroscopies (ADRESS) beamline \cite{Strokov} at the Swiss Light
Source (SLS), Paul Scherrer Institut, Switzerland, using the
Super-Advanced X-ray Emission Spectrometer (SAXES)
\cite{GhiringhelliRSI}. A flux of $10^{13}$~photons$/$sec/0.01\%
bandwidth was focused to a spot size below $8\times100$~$\mu$m
(V$\times$H). RIXS spectra were recorded in typically 1 hour
acquisition time, achieving a statistics of 50-200 photons in peak
maxima. The combined energy resolution was 120~meV at the Cu $L_3$
edge ($\sim$930~eV). The Sr$_{14}$Cu$_{24}$O$_{41}$ single crystal
was grown with the traveling-solvent floating zone method. Samples
were cleaved ex-situ, producing a mirror-like surface with ${\bm
b}$-orientation. The sample was mounted with ${\bm b}$- and ${\bm
c}$-direction (leg-direction) in the scattering plane, as depicted
in the sketch of the experimental geometry in Fig.~1. Two geometries
were used with the angle between the incident (${\bm k}$) and
scattered (${\bm k'}$) light being 90$^\circ$ (upper part) and
130$^\circ$ (lower part). With this setup one can cover at the Cu
$L_3$-edge up to 90\% of the BZ in Sr$_{14}$Cu$_{24}$O$_{41}$ (the
lattice constant of the ladder system is $c_L=3.93$~\AA). Incident
light was linearly polarized either out of the scattering plane
(${\bm \sigma}$-polarization) or in the plane (${\bm
\pi}$-polarization).

The left panel in Fig.~1 displays Cu $L_3$ RIXS spectra of
Sr$_{14}$Cu$_{24}$O$_{41}$ measured at room temperature (RT) and at
15~K. Spectra were obtained with $\sigma$-polarized light at
20$^\circ$ grazing incidence in 90$^\circ$ geometry. The excitation
energy was detuned by $\sim0.2$~eV from the resonance maximum to
reduce the elastic contribution, indicated in the x-ray absorption
data (XAS) in the inset (acquired in TFY mode with a photodiode).
Both RIXS spectra reveal two intense well-separated structures. One
peak at zero energy loss consists of the elastic signal with
unresolved low-energy contributions from phonons and presumably
magnetic excitations of the chains. The second peak at final energy
loss represents a low-energy excitation. The position at around
270~meV corresponds to the energy range of intra-ladder exchange
coupling. Previous Cu $K$ RIXS investigations reported on charge
excitations in the energy range of 2-6~eV
\cite{Ishi2007,Higashiya2008}. The spectrum at RT is slightly
broader than at 15~K, which is the expected temperature dependence
of spin excitations.

\begin{figure}
\includegraphics[scale=1,clip,bb=99 72 510 300,width=5.8cm]{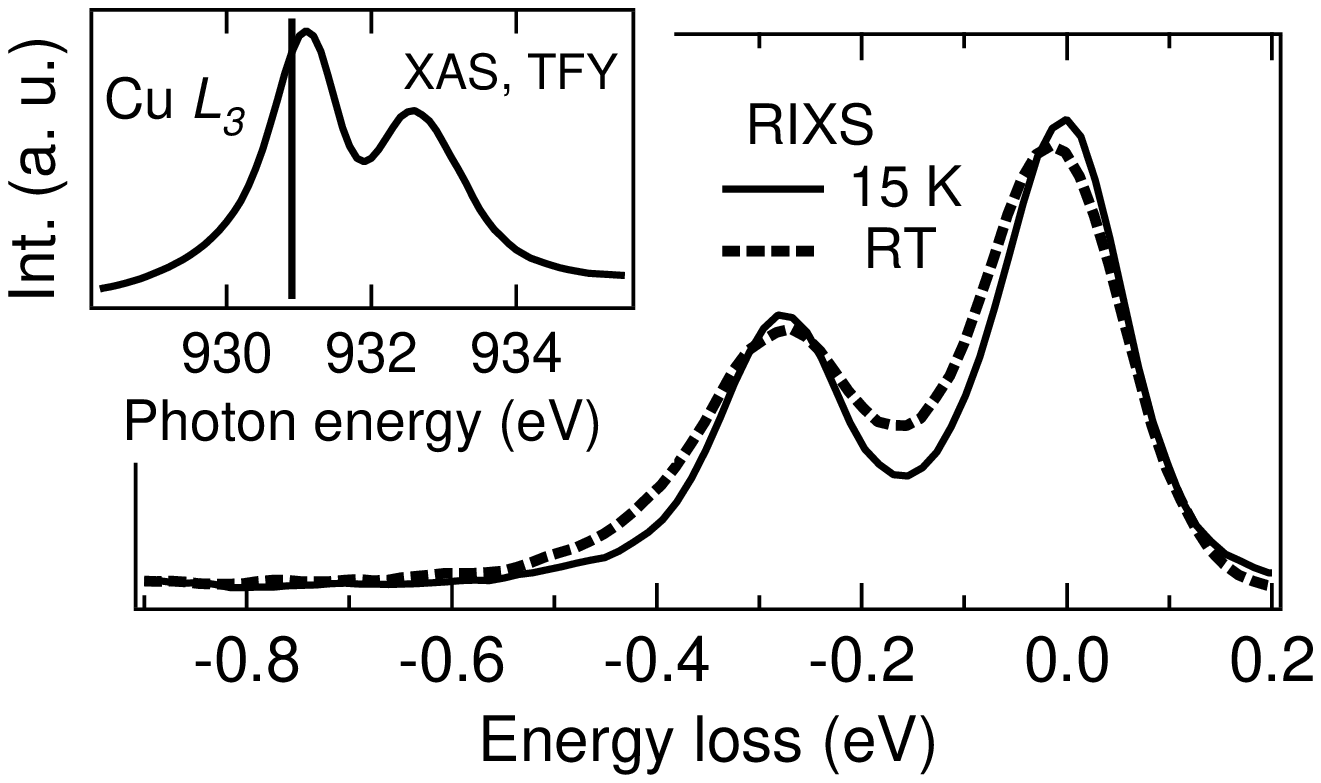}
\includegraphics[scale=1,clip,bb=157 402 342 783,width=1.65cm]{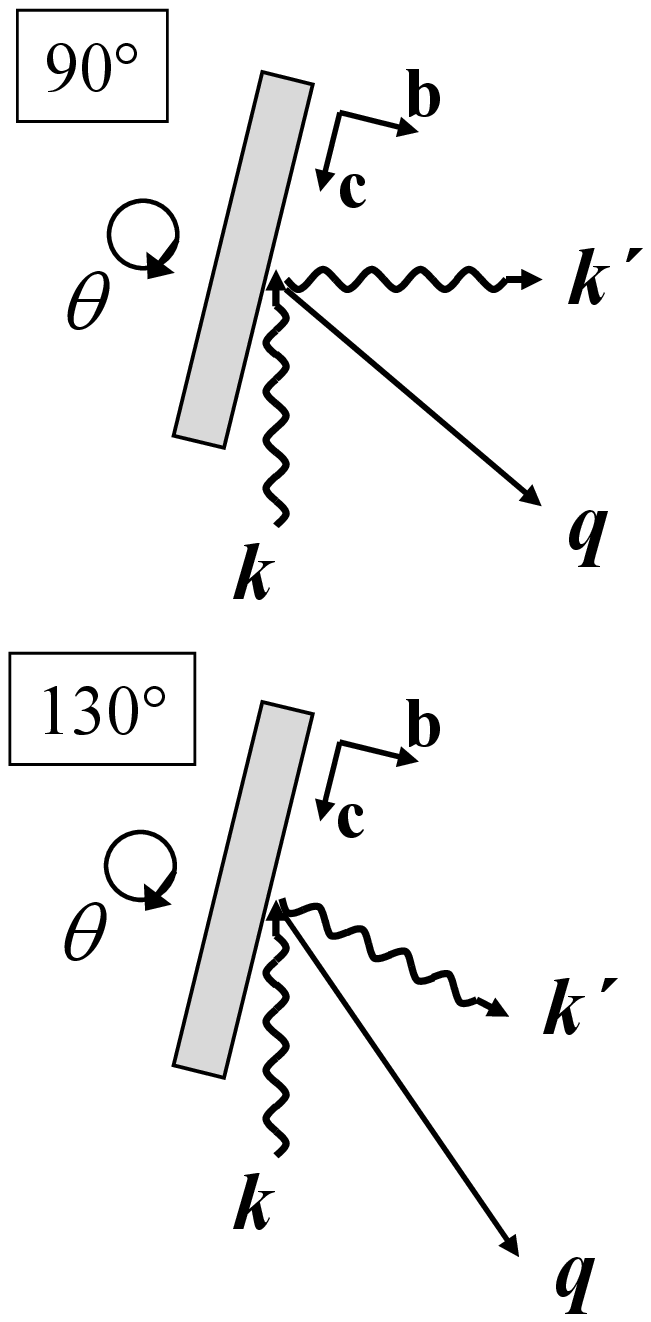}
\caption{Left: Cu $L_{3}$ RIXS of Sr$_{14}$Cu$_{24}$O$_{41}$
measured at 15~K (solid line) and RT (dashed line): 20$^\circ$
grazing incidence, 90$^\circ$ scattering angle, $\sigma$-polarized
light ($E \parallel a$). Excitation energy is indicated in the XAS
data (inset) in total fluorescence yield (TFY) mode, acquired at
15~K in the same geometry. Right: sketch of the experimental set-up:
90$^\circ$- (top) and 130$^\circ$- (bottom) scattering geometries,
${\bm k}$ (${\bm k'}$) denotes the wavevector of the incident
(scattered) light and ${\bm q}={\bm k'}-{\bm k}$ the transferred
momentum.}
\end{figure}

To understand the local vs. collective character of this excitation
we studied its dispersion upon $q_c$, momentum transfer along the
leg-direction. Since Sr$_{14}$Cu$_{24}$O$_{41}$ is a low-dimensional
system, where we expect no dispersion along ${\bm b}$, we could map
out $q_c$ by simply rotating the sample around ${\bm a}$. Momentum
transfer dispersion was measured at 15~K using the same photon
energy and polarization as in Fig.~1. RIXS data for different $q_c$
transfer are presented in Fig.~2. The upper panel displays raw
spectra normalized with acquisition time and a geometry-dependent
factor accounting for variations in scattering volume
\cite{Guinier1994}. Spectra for $|q_c|<0.21\times2\pi/c_L$ were
obtained in 90$^\circ$ geometry (black) and for higher momentum
transfer in 130$^\circ$ geometry (gray). All spectra reveal the same
pronounced magnetic mode as in Fig.~1, dispersing strongly across
the BZ \cite{elastic}.

\begin{figure}
\includegraphics[scale=1,clip,bb=19 0 387 374,width=7.4cm]{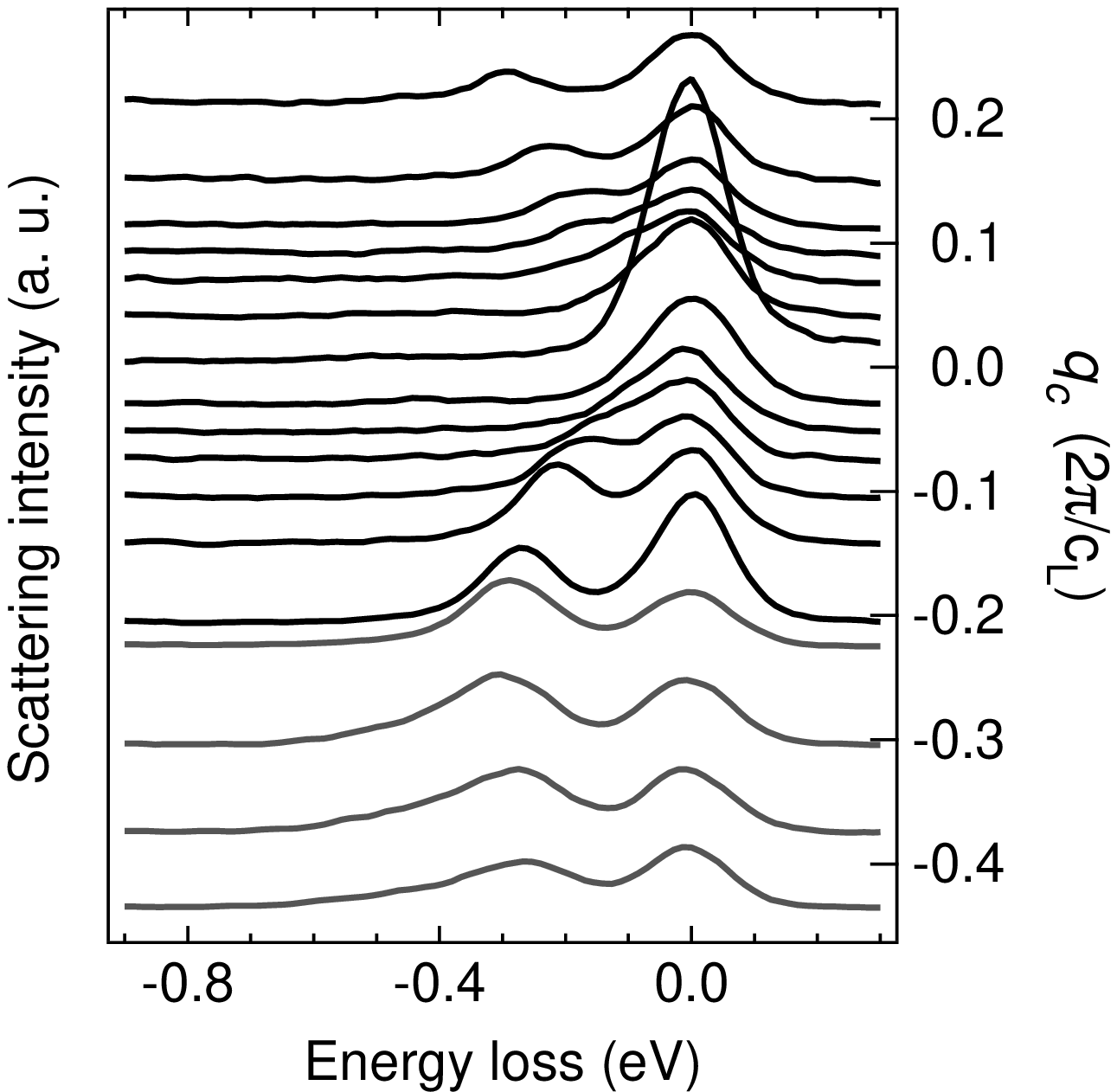}
\includegraphics[scale=1,clip,bb=27 4 412 255,width=7.4cm]{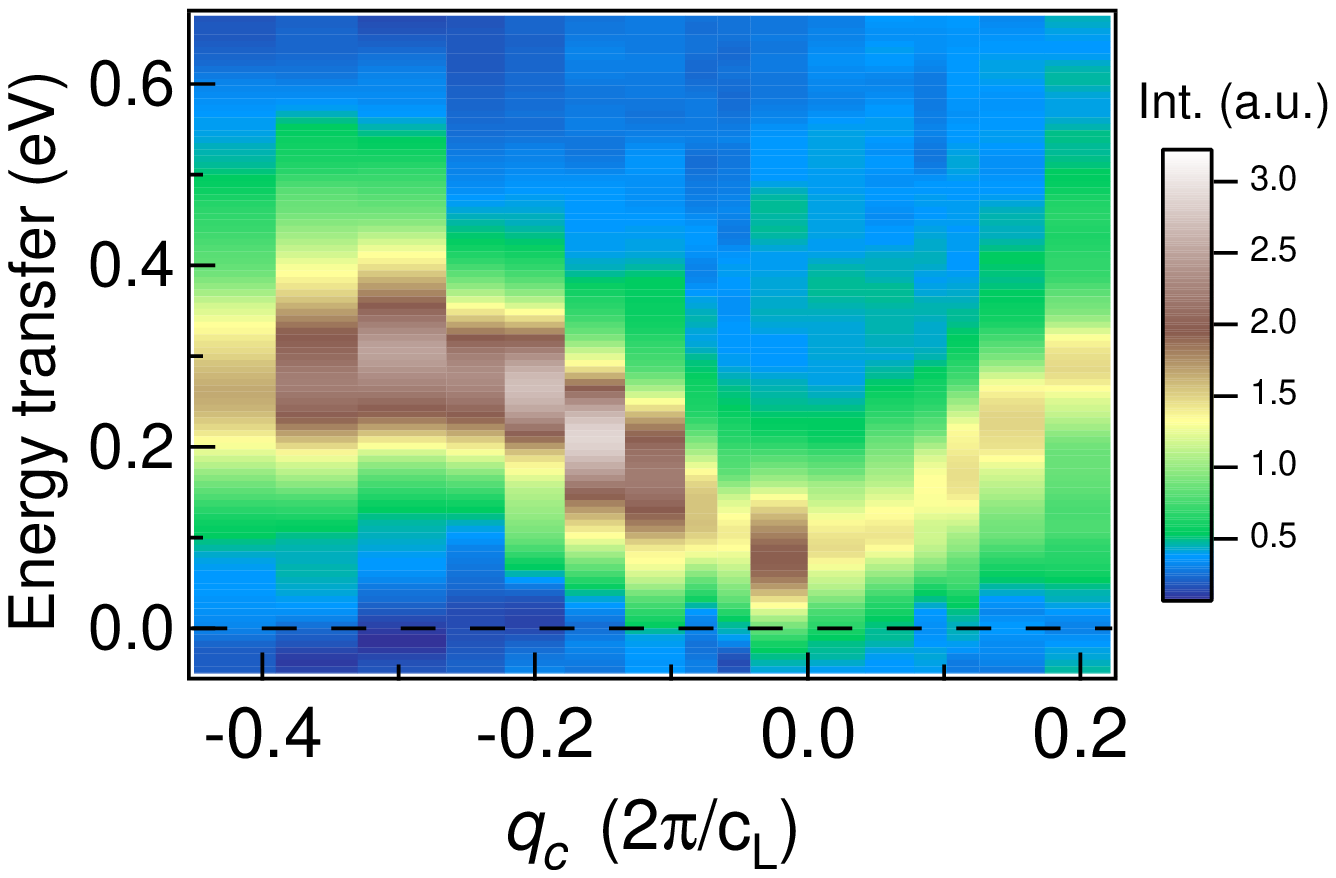}
\caption{(color online) Dispersion of magnetic excitations in Cu
$L_{3}$ RIXS from Sr$_{14}$Cu$_{24}$O$_{41}$. Upper panel:
experimental spectra obtained at 15~K with $\sigma$-polarized light.
Lower: RIXS data as intensity map vs. momentum and energy transfer
after subtraction of the elastic signal.}
\end{figure}

The lower panel of Fig.~2 displays an intensity map of the RIXS data
plotted vs. momentum and energy transfer. The elastic contribution
has been subtracted, after fitting each experimental spectrum with
two Gaussians. The magnetic excitation is seen here to disperse
around the BZ center ($q_c=0$), where it also reaches its minimum in
energy loss. With larger $|q_c|$ it moves first towards higher
energy losses and is then folding back close to
$q_c=0.3\times2\pi/c_L$ towards the BZ edge. The width increases
slightly towards BZ edge.

This dispersing behavior reveals the collective character of the
observed magnetic mode. Similar dispersion has been partially
observed in INS from La$_4$Sr$_{10}$Cu$_{24}$O$_{41}$
\cite{Notbohm2007}. Our dispersion curve revealed by the Cu $L_3$
RIXS data matches well with the two-triplon mode measured with INS,
however, the observed intensity vs. momentum dependence is
different. While INS intensities are high in the region $q_c>0.25$
and low for small momentum transfer, our RIXS data reveals uniform
intensity of the excitation over the BZ. The magnetic excitation can
be detected close to the BZ center, where it approaches a finite
energy loss value.

\begin{figure}
\includegraphics[scale=1,clip,bb=10 0 521 333,width=7.5cm]{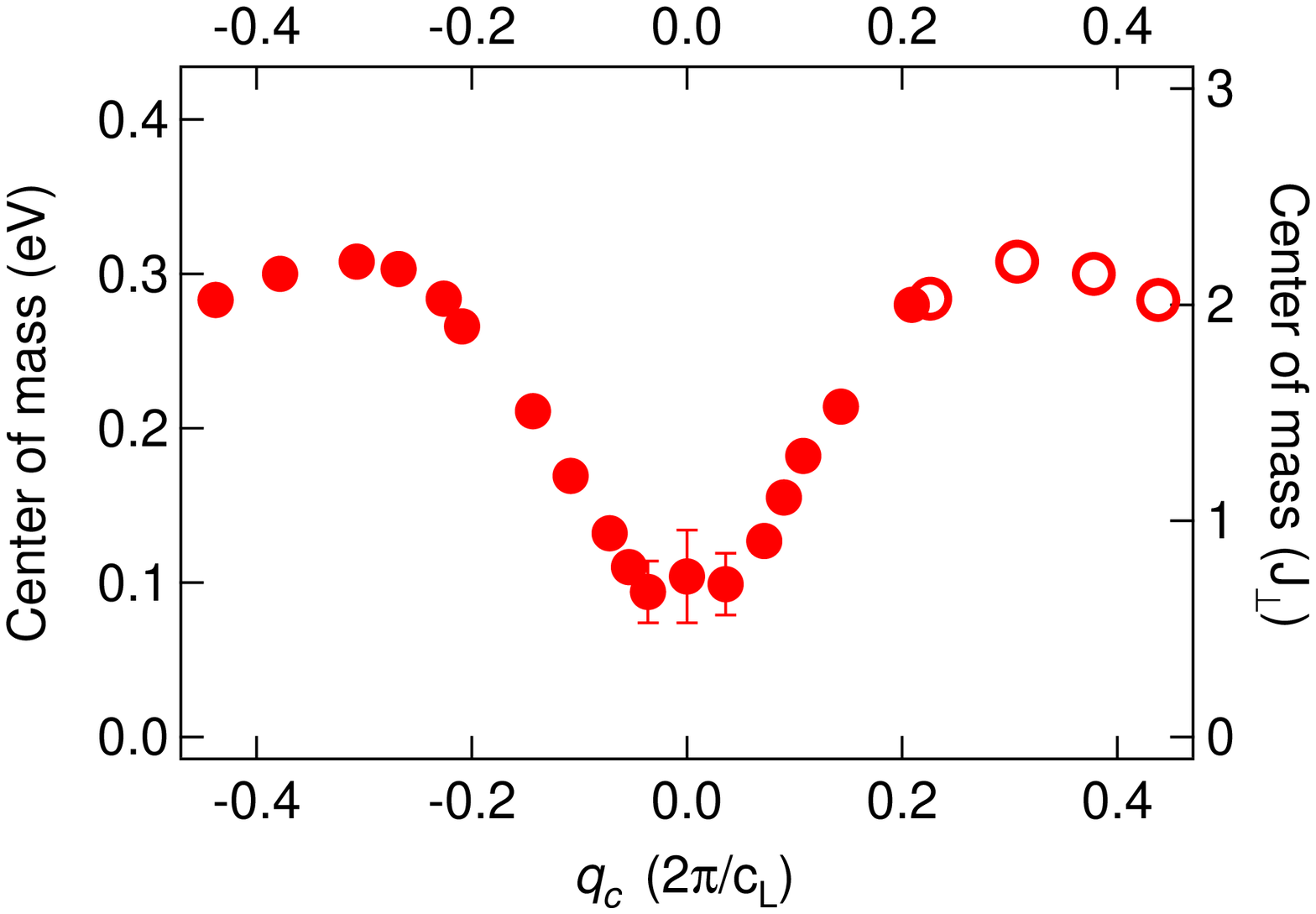}
\includegraphics[scale=1,clip,bb=-15 0 490 200,width=7.5cm]{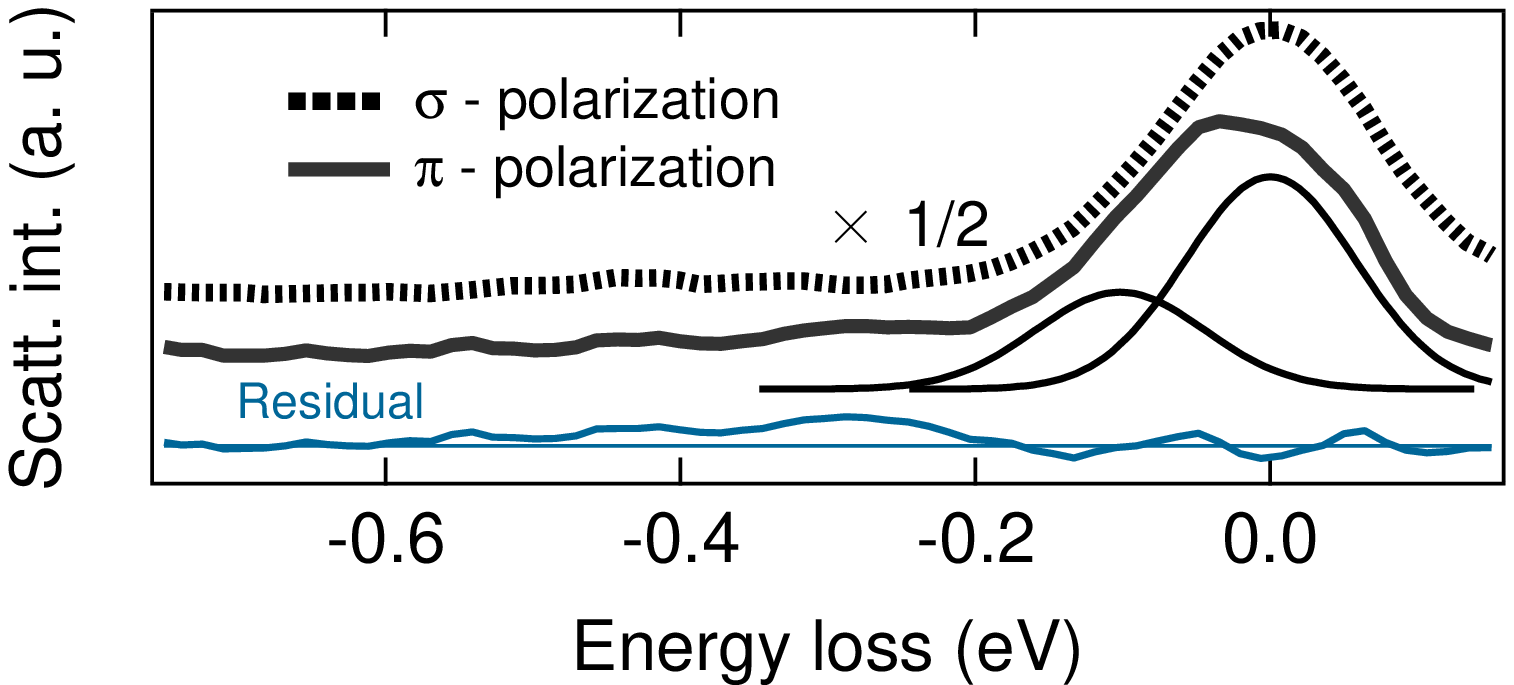}
\caption{(color online) Upper panel: dispersion curve of the
collective spin excitations across the first BZ. The open symbols
represent mirrored data points from 130$^\circ$ geometry. The right
axis is scaled in units of $J$ along the rungs ($J_{\perp}$),
extracted from the theoretical model below. Lower: RIXS spectra
measured close to the $\Gamma$ point for $q_c=-0.04\times 2\pi
/c_{L}$, using $\sigma$- and $\pi$-polarized light. The
$\pi$-polarized spectrum is fitted by two Gaussians, the residual is
represented by the thin blue line.}
\end{figure}

Using the procedure described above, we extracted from our data the
center of mass of the magnetic excitation for the different points
in the BZ. The resulting dispersion curve is presented in upper
graph of Fig.~3. Data points corresponding to spectra measured in
130$^\circ$ scattering geometry have been mirrored upon the BZ
center. Values close to $q_c=0$ ($q_c < 0.05\times 2\pi /c_{L}$)
were obtained from $\pi$-polarized data to suppress contribution
from the elastic channel. We can follow for the observed energy gap
the energy loss of $100\pm30$~meV and for the maximum loss value
320~meV. The lower panel of Fig.~3 presents fit of a $\pi$-polarized
spectrum (thick solid line). Interestingly, the residual reveals
weak intensity in the energy range between -0.2 and -0.6~eV.

To assign the magnetic excitation which we observe in our RIXS data,
we take a look at the electrical dipole transition selection rules.
We confine our considerations to collective excitations and neglect
in a first approximation the spin-orbit coupling \cite{remark}. As a
consequence L and S remain good quantum numbers and in the electric
dipole approximation only transitions with $\Delta L=\pm 1$ and
$\Delta S=0$ will be allowed. In the present system the elementary
magnetic excitation ({\it triplon}) consists in promoting a
spin-singlet into a triplet, which clearly leads to $\Delta S=1$ and
not to a dipole-allowed transition. Having a $\Delta S=0$ excitation
necessarily means exciting an even number of these triplons
together, the leading process being thus a two-triplon excitation.
We simulate which kind of magnetic excitations will occur in the
ladder system in the Cu $L_3$ RIXS process by confining our
considerations to the optical selection rules. As a minimal model an
effective Hubbard Hamiltonian downfolded from a multi-band Hubbard
model is used \cite{ZhangRice1988,Eskes1991}:
\begin{equation}\label{hamilt}
{\mathcal H}=\sum_{\langle i,j\rangle ,\sigma}t_{ij}
\left(d^\dagger_{i,\sigma}d_{j,\sigma} +\ {\rm h.\ c.}\right) + U
\sum_{i} n_{i,\uparrow}n_{i,\downarrow}
\end{equation}
with $n_{i,\sigma}=d^\dagger_{i,\sigma}d_{i,\sigma}$. The hopping
parameters in (\ref{hamilt}) are taken as $t_{\perp}=0.35$~eV,
$t_{\parallel}=0.3$~eV while the on-site Coulomb repulsion is
$U=3.5$~eV. According to results of XAS, the concentration of holes
in the ladder system is smaller than $10\%$ \cite{Nuecker2000}. We
consider therefore that we are at half-filling (1 hole per Cu-site).
From the above parameters we extract $J_{\perp} \sim 140$~meV, which
is close to the experimental value measured with INS or Raman
scattering \cite{Schmidt2005,Notbohm2007,Gozar2001}. In this picture
the experiment can be considered as a coherent process of two
optical transitions: promoting a Cu-$2p$ electron to the $3d$-band
and the recombination of the $2p$-hole with an electron from the
$3d$ band. In essence, this can be rationalized as having a
non-magnetic impurity in the $3d$-band for the intermediate state.
In the lower energy-loss region, this will naturally lead to
magnetic rearrangements and finite overlaps with final excited
states in different symmetry sectors than the ground-state. The
Hamiltonian in Eq.~(\ref{hamilt}) was fully diagonalized for an
8-site cluster and eigenvalues and eigenvectors were obtained for
the ground and final states at half-filling (8 particles) and for
the intermediate states (7 particles). Spectral intensities were
calculated using the Kramers-Heisenberg formula \cite{Kotani2001}
with the optical transition operator expressed in the hole
representation: ${\mathcal O}_{\bm k}=\sum_{j,\sigma}
p^{\dagger}_{j,\sigma}d_{j,\sigma}e^{i{\bm k}\cdot{\bm r}_j}$, $d$
removing a hole in the $3d$-band and $p^\dagger$ creating one in the
Cu-$2p$ shell. Presence of a Cu-$2p$ core-hole in the intermediate
states was accounted for by an on-site Coulomb interaction
\cite{hardrixs}. The calculated RIXS profiles for the accessible
$k$-points are displayed in Fig.~4. These spectra show a dispersive
low-energy excitation of energy loss $\leq400$~meV. Comparison of
the energy position in the simulated and the experimental data
reveals an offset of $\sim100$~meV between them, which can be
ascribed to finite-size effects. Nevertheless, despite the finite
cluster-size, the excitation disperses in qualitatively the same way
as in the experiment. We therefore conclude that the observed mode
in our RIXS data is in the $\Delta S=0$ channel and that the main
contributions are two-triplon excitations in the ladder subsystem.
The observed energy gap of $100\pm30$~meV in our our experimental
data is attributed to the two-triplon energy gap.

These results are in contrast to RIXS observations from a
3-dimensional antiferromagnet NiO, where it was found that the main
contribution to magnetic excitations are in the local spin-flip
channel \cite{Ghiringhelli2008}. On the other hand, our
interpretation is inline with observations of low-energy excitation
spectra from IR spectroscopy, INS and with spectral-density
calculations on cuprate ladder-systems
\cite{Windt2001,Sugai2001,Schmidt2005_2,Notbohm2007,Schmidt2005}.
Comparing the data with spectral density calculations for
multi-triplon contributions by Schmidt and Uhrig \cite{Schmidt2005}
indicates that the dominating magnetic mode observed in our Cu $L_3$
RIXS data corresponds to the lower boundary of the two-triplon
continuum.

\begin{figure}
\includegraphics[scale=1,clip,bb=18 58 706 522,width=6.5cm]{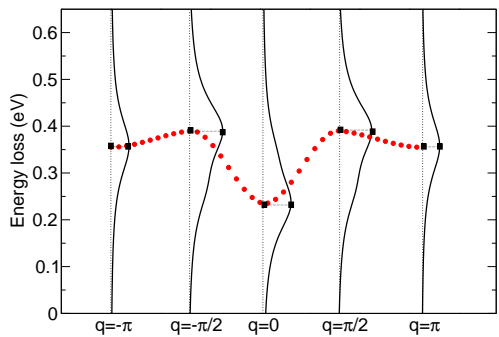}
\caption{(color online) Cu $L_3$ RIXS simulation for an effective
Hubbard model on an 8-site ladder cluster.}
\end{figure}

To summarize, we have investigated the two-leg spin ladder compound
Sr$_{14}$Cu$_{24}$O$_{41}$ using RIXS at the Cu $L_3$ edge. Our data
reveal that the dominant signal in the RIXS process in this system
is due to two-triplon excitations. Therefore we prove that this
technique is able to probe purely quantum mechanical fluctuations,
in addition to spin-wave excitations in a long-range ordered magnet.
The experimental results are supported by simulations based on
optical selection rules and an effective Hubbard model for a
finite-size cluster. Uniform RIXS cross-section over the BZ allows
us to trace these collective modes down to zero momentum transfer,
where a two-triplon spin gap of $100\pm30$~meV is found. We
demonstrate that RIXS is emerging as a powerful probe of magnetic
excitations, complementary to INS with respect to accessible energy
and momentum transfer.
\\

This work was performed at the ADRESS beamline of the SLS (Paul
Scherrer Institut) using the SAXES spectrometer developed jointly by
Politecnico de Milano, SLS and EPFL. We gratefully acknowledge M.
Kropf and J. Krempasky for their technical support, M. Grioni and C.
Dallera for their contribution to commissioning of the SAXES
spectrometer and C. Quitman for his critical reading of the
manuscript. Work at the EPFL is supported by the Swiss NSF and work
at CEA-Grenoble by the Indo-French project 3408-4.



\begin{thebibliography}{99}
\bibitem{remark}Taking explicitely the spin-orbit coupling of the Cu-$2p$
shell into account would allow for weak local spin-flips.
\end{thebibliography}
\end{document}